# Prospects of Solving Grand Challenge Problems

Rajan Gupta*

T-8, MS-B285, Los Alamos National Laboratory, Los Alamos, NM 87545

The recent woes of the supercomputer industry and changes in federal funding have caused some scientists to re-evaluate the means by which they hope to solve Grand Challenge problems. I evaluate the potential of Massively Parallel Processors (MPP) within this context and the state of today's MPP. I stress that for solving large-scale problems MPP are crucial and that it is essential to seek a balance between CPU performance, memory access time, inter-node communications, and I/O. To achieve this it is important to preserve certain characteristics of the hardware while selecting the hottest processor to design the machine around. I emphasize that for long term stability and growth of parallel computing priority should be given to standardizing software so that the same code can run on different platforms and on machines ranging from clusters of workstations to MPP.

16 Oct 1994

# 1. Introduction

The premise that millions of simple off-the-shelf computers working together can provide many-fold increase in computational power over conventional supercomputers and be more cost effective has been the cornerstone of parallel computing since its conception. This maxim is as true today as it was in the early eighties. The challenges of MPP have been in the development of efficient inter-node communication networks, fast memory access, and the software paradigms needed to program these machines efficiently. While the industry has made tremendous strides in both hardware and software, massively parallel machines are still not the obvious choice for large-scale applications. As a scientist who is interested in solving problems using MPP and needs a factor of a thousand to a million times the computer power available today, I present my views on what it would take to get there.

There are five points I would like to make. In Section 2 I present, as an example of successful code optimization, our strategy for lattice Quantum Chromodynamics (LQCD) calculations on the CM-5. The existing MPP, CM-5, Cray T3D, and Intel Paragon, are compared in Section 3. I show that the performance of all current MPP is limited by memory access time. In order to overcome this bottleneck in future MPP I suggest that the CPU and memory design should support pipelined memory access and have register files of a minimum size, which my experience suggests should be at least 64 words long. The sociological reasons why MPP have not been a resounding success are given is Section 4. The advantages and disadvantages of MPP versus clusters of workstations are discussed in Section 5. I argue that large-scale fine-grained problems require the resources of an MPP, while cluster of workstations are best suited for coarse-grained problems, and I make a case for the dire need for common software that can run on a variety of platforms, and on environments ranging from clusters of workstations to MPP. In Section 6 I describe some of the ongoing LQCD motivated projects to construct specialized MPP in the teraflop class. These special purpose computers (defined to be those built specifically to solve a limited class of problems) can have a significant impact on Grand Challenges problems and help in the development of MPP technology for the next few years. Finally, the conclusions are presented in Section 7.

# 2. Road map to optimization on the CM5

I give only a simplified view of LQCD calculations since this is a diverse audience. The core operation that constitutes the bulk ($> 90\%$) of computer time is a matrix multiply that can be done in parallel at each site (or at least on every red or black site) of the lattice

$$A \; = \; B + C * D$$

where $A$, $B$, $C$, $D$ are $3 \times 3$ complex matrices. The matrix $D$ has to be communicated from a nearest neighbor in every $1/4 - 3/4$ of the operations. The geometry of the lattice is a 4-dimensional hypercube with periodic boundary conditions. Today we can simulate $32^3 \times 64$ lattices. However, looking into the future, the systems we would like to simulate are $100^3 \times 200$, *i.e.* $\approx 10^8 - 10^9$ points. The number of variables needed per point are approximately $100 - 200$. At each point the basic matrix operation has to be repeated $10^8 - 10^9$ times when working with what is called the quenched approximation, and $10^{12} - 10^{13}$ times with the full theory. This makes LQCD an extremely large data and memory intensive problem that will require (short of miraculous improvements in algorithms) petaflop years of computational power to solve completely. Even with these conservative estimates I believe that LQCD will be within the scope of MPP built using mass-produced hardware that will be available in the next decade or two.

LQCD is an extremely simple problem to parallelize. It requires only nearest-neighbor communications along the grid axes and the computation to communication ratio is relatively high. For every 24 bytes communicated, one typically performs 66 floating point operations. In addition, all the code is user-developed. Past experience shows that it takes on the order of ten to twelve man-months to develop highly tuned codes for a new architecture machine. For these reasons LQCD calculations are always amongst the earliest users of MPP and have provided significant guidance in developing and debugging both the hardware and the software.

Given the characteristics of LQCD, we chose to use the data parallel programming model on the CM-5 as it is simpler and therefore faster to implement than message passing. Data parallel programming has the potential drawback that the user makes no distinction between points in the interior and those on the boundary of the sublattice stored in each processor. Thus, there are unnecessary memory moves when the nearest-neighbors of interior points are addressed. In a message passing program one can circumvent this by having separate loops over interior and boundary points, but with some overhead due to the specialized array layouts. On the CM-5 it turns out that there is no difference in efficiency between data parallel versus message passing versions of LQCD codes. The advantage of message passing programs, as demonstrated by the MILC collaboration [1], is portability. Their code, initially developed for the Intel Paragon, ported successfully to the CM-5 and the Cray T3D with a few day's work.

Our program is written in CM Fortran with all the computationally intensive portions isolated into short subroutines. We choose CMF because (a) on the CM-5 it was the least buggy and most optimized compiler, (b) it was the most convinient language for my collaboration, and (c) it is adequate for our purposes as we aimed to get performance by replacing the compute intensive subroutines by CDPEAC (assembly language) versions. Keeping the overall control in CMF has allowed us to develop codes for new physics quickly and to incorporate algorithm changes with very little effort.

The routines that we converted to CDPEAC are short, typically 10 to 100 lines of code. We maintain both the CMF and CDPEAC versions of these modules and are thus able to make performance and integrity checks at any given time. The CDPEAC routines allow us to choose the vector length appropriate to each case, optimize loads and stores, and merge loads with flops efficiently. We find it efficient to load vectors as needed and do not pay any penalty compared to an optimization strategy in which a given vector is loaded only once and preserved in the registers until all the operations involving it are complete. The advantage of our approach is that the code and array layouts are much simpler and therefore easier to debug and modify.

As a result of CDPEAC optimization, the compute intensive modules achieved $40-50$

Megaflops/node on single adds or multiplies and $90 - 100$ Megaflops/node on chained multiply-adds. The 25% loss in single node performance is due to the memory structure on the CM-5, *i.e.* DRAM page faults reduce the peak performance from 128 to 100 Megaflops. Internode communications were implemented through CSHIFT as we were unable to improve upon this TMC supplied communications routine. Communications reduced the overall sustained performance of our codes to between 25 and 35 Megaflops/node. Another place where we were not able to come up with a reasonable optimization strategy is the case of masked operations. Masked operations in CMF are implemented only at the time of the final store, thus a red-black masking reduces the performance by a factor of roughly two.

The entire process of code development (50,000 lines of CMF) and CDPEAC optimization took us ten man months. We developed and optimized the code during the acceptance test period and were in full production mode by June 1993 when the CM-5 at LANL was first made available for long runs. At that point our code was $\approx 4.5$ times faster than the CM Fortran version. Significant compiler improvements brought this factor down to $\approx 2$ by Dec 1993, and there have not been any appreciable changes since.

To summarize, our optimization strategy on the CM-5 allowed us to get the best possible performance from day one and decoupled us from the day to day changes of the CMF compiler. We had the freedom to stay with a given working version of the compiler as we did not rely on it for performance. In addition to floating point performance, the CM-5 has fast disk access to the SDA and a reasonable I/O bandwidth that allows us to write data to an Exabyte tape unit. This entire package (floating point performance, SDA, and tape storage) is necessary in order to successfully undertake our Grand Challenge calculations. Looking beyond our interests alone, we feel that Thinking Machines Corporation has developed a high performance MPP that provides a stable production environment for a large class of Grand Challenge problems. Their collapse, therefore, is a tragedy for the nation as they were both the pioneers and leaders in MPP technology.

Since this is a workshop on debugging and performance tools, it is perhaps mandatory that I address the question -- what tools would have made our task easier? There are two things we would have liked, first that PRISM (the X-window based debugger) had been available during the code development stage (Jan-May, 1993), and second a better array of hardware diagnostic routines to identify chips that fail intermittently or under certain circumstances only. Other than CDPEAC optimization, we do not use any fancy constructs or try to push the limits of the CMF compiler. As a result we have needed the minimum in terms of debugging and performance analysis tools. Prism is a remarkable debugging tool and, in the last year, we have often wondered why it is not available for UNIX workstations like SUN's?

## 3. Comparison of existing MPP and the lessons learned

The three MPP that I am familiar with are the CM-5, Intel Paragon, and the Cray T3D. Each of these uses a different custom-designed network and a different topology of interconnections between the nodes. In order to compare them, I give a brief description of their main features.

The CM-5 uses a custom designed floating point accelerator (DASH chip) that also serves as the memory controller. Each node is controlled by a SPARC chip and the network is a fat tree. The communication bandwidth is realistically 10 Mbytes/sec and is the major bottleneck in the CM-5. The network interface (NI) and the network are proprietary while the memory and disks in the SDA are commercial commodities. The peak I/O bandwidth to the SDA is about 100 Megabytes/sec and our codes sustain a significant fraction of it unless there is competition from other jobs. The peak speed per node is 128 Megaflops while the LQCD codes sustain $\approx 30$, the main limitation being internode communications.

The T3D is connected as a 3-dimensional toroidal grid with a data bandwidth of 300 Megabytes/sec in each direction. Each node is controlled by a DEC Alpha chip with a peak speed of 150 Megaflops. The host processor is a YMP or a C-90; the attached disks are Cray DD301 which can sustain a transfer rate of 50-100 Megabytes/sec depending on the configuration. The LQCD kernel sustains $\approx 60$ Megaflops on a single node with data restricted to fit in the cache, but only $\approx 25$ Megaflops when arrays were scaled up to realistic sizes. There is no chaining of adds and multiplies on the Alpha and for performance one needs to unroll the matrix multiply loops and separate multiply and add operations into two separate loops. Since the processor is 6 deep (it takes 6 cycles to get back the result), it is therefore difficult to optimize computations with only a 32 word register file. A positive feature is that the fast communications network makes the latency of on-node versus off-node memory access virtually identical. The bottom line is that, for large-scale problems, the potential of the FPU is not realized due to the slow memory access.

The T3D is built out of mostly custom designed chips. The DEC Alpha is expensive and its power consumption is too high for large-scale packaging. Given that the performance of the T3D is limited by memory access time, I believe that it may serve CRI the interim purpose of getting a share of the market (largely based on the past reputation of their X-MP, Y-MP, and C-90 line) while developing the first round of parallel software, but eventually they will need to move to mass produced technology if they hope to provide a cost-effective product in the teraflop range.

The Intel Paragon is arranged as a 2-dimensional mesh with a data bandwidth of 200 MB/sec/channel using a custom designed NI chip. The processor is the i860XP with a peak speed of 100 Megaflops in single precision mode. The memory access is slow if data is not in cache or lies outside the 512 word page limit as there is a DRAM latency of 7-8 cycles for page faults. LQCD codes sustain $\approx 27$ Megaflops on a single node [1]. Internode communications are fast, and the performance stabilizes at $\approx 23$ Megaflops for 16 or more processors. The positive feature of the Intel i860 line has been that the codes are compatible between the hypercube, Delta, and Paragon. The critical drawback, so far, for large-scale calculations is that the I/O bandwidth to the outside world is very low.

Thus one finds that despite their completely different hardware and peak performance of the FPU, all three MPP give roughly the same per node performance. The reason for this is that they are all limited by memory access or inter-node communications. The floating point unit can, in each case, process data much faster than the network can supply it. Looking at the last 10 years (see the chart in Greg Papadopolous's talk at this meeting), it is clear that the increase in performance of the FPU has outstripped that of memory access and I/O, and this trend will continue. Thus, the emphasis for future hardware development should be on getting data to the registers efficiently rather than just on the peak rate of the FPU.

There are three features of the hardware that I believe are essential for getting performance. First, memory access needs a pipeline architecture. In most large-scale applications the efficiency of an intermediate cache based architecture is very low and requires significant hand tuning of the code. Therefore, my preference is for pipelined access of main memory rather than through a cache. Second, the length of the pipeline required to hide the memory latency should dictate the length of the register file. Our present experience suggests that this length is 8, in which case the register file should be at least 64 words long. Lastly, data from neighboring elements is needed only as part of some calculation. It should, therefore, go directly from the NI to the register file rather than through the processor's memory. I believe that imposing some standards with respect to these features will make the development of optimizing compilers possible and provide compatibility of software across platforms.

Another drawback is that each of these machines has a very large number of custom designed and proprietary chips. As a result they are neither very cost-effective nor massively scalable machines.

Finally, the most critical point is that each of the three has its own proprietary software. The cost of developing the software is very high and the lag-time between the delivery of the hardware and efficient implementation of user codes is very large. The typical lag time has been two years, while the lifetime of a given technology is roughly 3-4 years. On the Connection Machines, codes written in CMF port between the CM-200 and the CM-5 but the compilers are very different and routines optimized in the respective assembly languages are totally unportable. Intel has maintained the same software through their i860, Delta, and Paragon line, nevertheless they are still not able to deliver more than 25% efficiency for simple codes like LQCD. The delivery of the T3D started in late 1993 and its software environment is still evolving.

## 4. Why have the MPP not "succeeded"?

Why have the MPP not had as much commercial success as some of us had expected? The demands on MPP vendors have been formidable. They have had to develop all the hardware and software in house, and in addition teach the potential customers how to use the machines. This has required a large investment in money and talent. They have also had to deliver a much more powerful machine than the best supercomputers in order to make it worthwhile for users to switch. Lastly, since the MPP are supposed to be made up of mass produced chips, people expect their performance to keep pace with the latest developments in technology. These demands would not have been fatal if the revenue base for large scale computing had grown.

To understand why this revenue base did not grow as expected, we need to understand the sociology of industrial, research, and commercial applications that "need" many orders of magnitude more computer power than that provided by workstations but have been slow to adapt to MPP, or worse still, have ignored them altogether. To present my perspective I would like to partition the use of computers for such large scale projects into three categories –– evolutionary, revolutionary, and research.

Evolutionary projects are those for which computations have been done for the last forty years on whatever machine was affordable. The problem has been incrementally scaled up in size as the computers became more powerful. The software base is, therefore, very large as it has been developed over a very long period of time and, unfortunately, all of it is for serial or, at best, for vector machines. The human investment needed to port these hundreds of millions of lines of code to MPP is very large and, given the mode in which this type of computing has functioned, there is very little intellectual or financial pressure to migrate to MPP. A very large proportion of industrial applications fall into this category.

Revolutionary projects require that the old way of doing business be discarded, as for example when a completely new product or process is needed. Also, the simulations needed to design this product require many orders of magnitude more computer power than was previously available to the R&D group, necessitating new ideas and algorithms.

A prime example of the revolutionary use of computing is computer assisted animations by the entertainment industry. The hardware best suited for this application seems to be a cluster of multi-processor Silicon Graphics machines. Since computer assisted animations became fashionable almost at the same time as parallel computing, the software development effort has been, to a large extent, for parallel processing. I give more examples of industrial problems that require large-scale computing in Section 5 where I compare clusters of workstations and MPP.

Research projects are, loosely speaking, those being addressed by scientists at universities and national labs and involve new algorithms and their implementation on emerging computer technologies. One would expect MPP to have their initial impact here and that has indeed been true. These projects are largely funded by federal or state grants and the current climate of funding uncertainty has had a very negative impact. It is essential for the success of this technology, including its timely transfer to industry, that a sufficiently large pool of the state-of-the-art high performance machines be available to researchers for the development of the new generation of applications.

It is clear that the active participation of industry is needed, as universities and national labs do not provide a sufficiently large revenue base for three or even two MPP vendors to thrive. Unfortunately, the industrial world is, by and large, not revolutionary. For example, automobile manufacturers continue to produce essentially the same car year after year making only cosmetic changes. The managers and most of the scientists predate the MPP and are understandably not overly anxious to switch. Their reluctance to embrace a new technology is strengthened by the fact that all the MPP are significantly different, there is a very large installation cost, and there is no guarantee that a particular vendor will be around in a few years time. Also, the transfer of technology from academia to industry is proceeding at a rate that is very slow compared to the mean lifetime of a MPP. This is largely due to the long time required for the development, incorporating the latest algorithms, of a new generation of parallel codes for industrial applications. The situation should change significantly when the parallel codes being developed start producing significantly better data. Then the potential for revolutionary changes in design through simulations should become obvious. At that time industry will have no choice but to embrace and support this technology.

To summarize, a number of factors, including increased performance of workstations, lack of an easy migration path from existing serial code to MPP, a large financial overhead in the development of system software, lack of stability amongst the MPP vendors, and a climate of uncertain funding situations, have contributed to a slow acceptance of MPP as a vital technology. I feel that in the long run MPP, and more generally parallel computing, will dominate industrial design and production. It is therefore essential that MPP technology be adequately funded through the current development stage.

## 5. Closely coupled MPP versus large clusters of workstations

There is wide spread belief that MPP technology has yet to prove itself and that it is very expensive to convert codes developed incrementally over the last 40 years to MPP and, at the same time, there is a growing recognition that parallel computing is the way of the future. In the face of these conflicting beliefs, there has grown tremendous enthusiasm about the potential of large clusters of workstations. There are three reasons for this, (a) the power of individual workstations has grown by a factor of 100 in the last 10 years, (b) there is a very large user base and a large number of third party software vendors, (c) there is no large setup cost to adding more nodes, and (d) workstations connected via the internet (or LANs) are idle for a very large fraction of the time and can be tapped into for free during that time. So let me discuss this resource from the perspective of solving LQCD and other fine grained Grand Challenge problems.

Fundamentally, there is no conceptual difference between closely coupled MPP and a very large cluster of workstations with regards to parallel computing. I will assume the ideal situation that both types of systems are made up of similar mass-produced chips and communications are through message passing. It is then clear that any code that can run on a cluster can run efficiently on a MPP, while the converse in not true for all problems, especially Grand Challenges. The question of efficiency includes cost, communication latency, and turn around time between the formulation of the problem and getting the final results. When evaluating the two, the focus should be on the ability to sustain the production environment appropriate for solving real problems and not just in a proof of principle.

The technical problem in parallel computing common to both platforms is in developing the software (operating system and compiler) that allows (a) efficient implementation of codes that are intrinsically parallel, (b) a production environment for problems that require many months equivalent of computing resources and have a significant amount of I/O to and from external storage devices, and (c) the most transparent migration path for existing serial codes to these environments. It is important to accept the underlying similarity of the two platforms and develop software that is common to both. I stress in the strongest possible terms that this should be the national goal. Once we can run the same problem on either platform, then the question of efficiency of a given platform can easily be decided by trial on a case by case basis.

My contention is that the problems best suited to run on a cluster of workstations are the very coarse grained. I define these problems as ones that can run on a single (or few) workstation and produce the desired result in a reasonable amount of time, but have to be run thousands of times in order to gather statistics or explore a multi-dimensional parameter space. Each sub-problem is essentially independent of the other and generates only a small amount of information that needs to be saved at the end or communicated to the other processes. Examples are data base searches, exploration of parameter space, statistical analysis of systems with a few tens of thousand of points or particles such as in biological molecules. It may well be that such problems constitute the bulk of society's computational needs today. I am therefore very optimistic that there exists a large commercial base to fund the development of software for clusters of work-stations. So if we firmly commit to ensuring that this development effort is common to both clusters of workstations and MPP, then we have a cost-effective path to teraflop and petaflop computing.

Let me now present the characteristic of a number of problems that definitely need closely coupled MPP. First I will make the point using LQCD and then move on to problems of interest to industry.

LQCD: This is a statistical Monte Carlo problem. At present we are doing simulations using $32^3 \times 64$ lattices. For each data point we first generate 25 Gigabytes of data. The code for doing this requires 3 Gigabytes of memory and 100 Gigaflop hours. This data set is then analyzed by 10 different programs each using between 1-10 Gigabytes of memory and 10-25 Gigaflop hours. With 15% of the resources of a 1024 node CM-5 we can process 100 data points in a year. I am willing to make a wager that, in the next five years, a problem with similar characteristics, if simulated on a cluster of 256 workstations, will not yield 500 data points in one year. The only conditions on the wager are that the cluster is not used as a dedicated resource (otherwise an MPP is clearly more cost effective) and that the performance of each node is similar to what we see on the MPP today, *i.e.* less than or equal to 50 Megaflops.

Comprehensive Hydrocode for Automotive Design (CHAD) [2]: The goals of this project are to simulate combustion inside the cylinders, underhood cooling, external aerodynamics, and air conditioning. The code uses a fully unstructured 3D grid with $10^5 - 10^8$ elements and roughly 500 words of memory per element. The algorithm is based on full implicit integration. To simulate combustion inside the cylinders, each run requires $\sim 10^5$ flops per cycle and per element, where a run consists of about 500 cycles. Global communication is needed roughly 100 times in each cycle, *i.e.* it is a fine grained problem. A single run, corresponding to a fixed cylinder geometry and air/fuel ratio, with $10^5$ elements would take roughly 300 hours on a workstation with a sustained performance of 20 Megaflops. At present a sample of, say, 100 runs is used to answer a specific question about the working of a given cylinder. To design a new cylinder would take many orders of magnitude more computing power. The CHAD code is being designed and written for parallel machines. Thus it can be used for evolutionary (on a small cluster of workstations) or revolutionary (MPP) design development. Only time will tell how aggressively the auto industry will use such tools.

Casting of Large and Complex parts [3]: In many industrial processes it is necessary to model casting processes with geometries as large as a cubic meter while at the same time resolve structures on the scale of millimeters or less. The parallel algorithms being developed at LANL use a finite volume, semi-implicit computational fluid dynamics ap-proach. The basic scheme requires the implicit solution of several parabolic and elliptic partial differential equations per computational cycle. For stability these computations need to be done in double precision. The lattice is a 3-D unstructured grid with $10^6 - 10^7$ cells. The number of words of data per cell are $200-300$. Data communications are global because of the unstructured grid and the implicit scheme. Typical CPU requirements are $50 - 100$ microseconds per cell per cycle on the CM-5, and a given run simulates flow for a duration of many seconds, *i.e.* many thousands of timesteps.

Another example of a fine grained problem that uses computational fluid dynamics methods is Process Chemistry. To simulate processes like metal and oil refining and aluminum smelting requires resources very similar to those described in the previous two examples.

Global Ocean Modeling [4]: This code uses a 3D grid with roughly $1280 \times 1000 \times 20$ cells requiring 4.5 Gbytes of memory. Each time step covers 30 minutes of ocean dynamics. A run evolves the state of the ocean over one month and requires roughly 8000 Mflop hours. The communication of data between cells during a run is fine grained. The data archived is $\sim 50$ Gbytes per simulated year, and so far 30 years of ocean dynamics has been simulated. The spatial resolution is currently 31 km at the equator and they hope to reduce this to a few kilometers. To get to this characteristic scale of ocean dynamics requires a thousand fold increase in computing power over what is currently available.

Lastly, let me describe the study of the large-scale structure of the universe as an example of an N-body problem [5]. A typical large gravitational N-body simulation contains 20 million mutually interacting particles. Through the use of "fast" approximate methods called treecodes, a simulation of about 1000 timesteps can be completed in around 30 hours on a 512 node Intel Paragon. The memory required is around 100-300 bytes per particle, which translates to 6 Gbytes for a 20 million particle simulation. Since the force law is long-range, global communication is required among the processors. By careful coding of the algorithm, the non-local data can be reduced to as little as 10%. This still requires communication of roughly 1 Mbyte of data per timestep (in several thousand messages of $\sim 100$ bytes) to each processor, giving a required bisection bandwidth of several hundred Mbytes/sec.

To summarize, I believe that for the foreseeable future clusters of workstations will be the medium of choice for coarse grained problems while MPP will remain the more effective platform for very large, fine-grained problems such as the ones I just described that need a dedicated production environment. This debate should in fact be irrelevant in the long run if we invest in system software, compilers, and computing paradigms that apply to both MPP and clusters of workstations. In that case we will be ready to exploit the resources of both types of hardware and determine from experience which is more practical. Unfortunately, in the short term we are faced with a wide spread belief that industry is not going to deliver the next generation MPP (teraflop capability) in the next few years or, if it does then it will be prohibitively expensive and not very reliable. Under these grim prospects for the availability of a large cost-effective MPP, what is the future for solving such Grand Challenge problems? The answer, in some cases, seems to be specialized research projects. In the next section I will briefly describe the on-going efforts to develop special purpose machines to solve LQCD.

## 6. Special purpose MPP for LQCD

Lattice QCD simulations have historically played a major role in the development of parallel computing. So it is no surprise that there are three major on-going projects to deliver teraflop scale computing capability. In each case the principle motivation for these projects is still LQCD. The software effort is minimal; it essentially stops at whatever is needed to run LQCD problems. The fact that a number of other Grand Challenges problems can be solved on these machines is an added bonus. Below, I give a brief description of each.

The CP-PACS project is based at Tsukuba University, Japan [6]. This is a joint undertaking between computer scientists and physicists at Tsukuba University and Hitachi Ltd. The processor is custom designed, based on the super-scalar Hewlett-Packard PA-RISC 1.1, but with enhancements. They have introduced slide window registers and preload and poststore instructions in order to overcome memory latency. The memory (DRAM) is pipelined by multi-interleaved banks and a storage controller. The communication is a 3D hyper crossbar −− x-direction, y-direction, and z-direction crossbars −− so that a maximum of three switches are needed for any transfer. The message passing is through wormhole routing and there is an option for global synchronization. The full machine will consist of 1024 nodes (they may increase the number of nodes dependent on funding ) with a peak speed of 300Gflops, have 500 Gbytes of RAID5 disks, and is expected to be ready by mid 1996.

The APEmille project is based at Rome University, Italy [7]. This is a follow up on the Ape 100 project which is currently being marketed by Alenia Spatzio. The geometry of the APEmille is a 3D torus. The programming paradigm is SIMD complemented with a local addressing feature. Each processing board consists of 8 nodes arranged in a $2 \times 2 \times 2$ topology. By having 2 multipliers and 2 adders per node they hope to achieve 200Mflop performance with a 50 MHz clock. Each node will have 2-8 Mwords of memory and can directly address the memory of its 6 neighbors through a communication device. To achieve a high I/O bandwidth each set of 128 nodes is connected to disks and peripherals through a host workstation with a 100 Mbyte/sec transfer rate. The time frame for completion of this machine is 1998.

I would like to dwell on the lone U.S. entry, the 0.5 Teraflop project [8]. This project is based at Columbia University and is the most cost effective. It uses a really mass produced chip, the Texas Instruments digital signal processor (DSP) TMS320C31, for 32 bit floating point operations with a per unit price of $\sim$ \$50. The DSP comes with a C compiler, consumes only 1 Watt of power, and has peak performance of 50 Megaflops. The big drawback is that the DSP does not have an operating system, however one expects to see 64 bit DSPs with more functionality in the near future. Each node is laid out on a $2.7'' \times 1.7''$ card mounted on a mother board through standard SIMM connectors. It consists of three elements, the DSP, a node gate array, and 2 Megabytes of DRAM as shown in Fig. 1. The only custom chip is the node gate array which serves as the memory controller, the network switch, and has a specialized cache for pre-loads. The DSP has some direct memory access (DMA) capability. The total power consumption of the node is about 2 Watts. To get to teraflop scale requires 16K processors which are mounted on 256 boards, each with 64 processors. Thus, by my criteria, this machine is pretty close to an ideal MPP in terms of its simplicity, mass-produced components, and packaging. The machine can be ready by 1996 for a projected $3 million in development cost and hardware. The extreme simplicity of this design is made possible because this machine is expected to solve only one problem, nevertheless, if funded and successful, this project will demonstrate that a system with 16K processors can provide a reliable production environment with current technology.

In addition to these three ongoing projects there is a proposal for a Multidisciplinary Teraflops project centered at M.I.T. [9]. This is an outgrowth of an earlier LQCD project and a collaboration with Thinking Machines Corporation. The goal of the scientists was to concentrate on the design of a single element –– the embedded accelerator. The accelerator design would use commodity products and be more general purpose, *i.e.*, be efficient for a number of large-scale problems. Sixteen of these accelerators (each capable of sustaining 200 Megaflops) would be attached to a node on the follow-up machine to CM-5 and piggy-back on TMC's message passing software. The project would use the next generation Connection Machine network for communications, which was supposed to sustain 500 Mbytes/sec transfer rate per node. Since there is not going to be a follow-up machine to the CM-5, this project is stalled until they can find an alternate network that suits their purpose.

I believe that the characterization "special purpose" or "general purpose" and the debate surrounding it should not be made a central issue in the development of large scale computing. In a sense all machines are special purpose in that they do some problems much better than others. In serial machines one does not complain about this because the software is common and issue is only of efficiency. In the field of MPP, the industry is still searching for solutions to problems of hardware reliability, software paradigms, communications, and I/O speed in addition to those of efficiency. In this semi-research environment we should learn what works from every successful project and move on.

If any or all of the above projects deliver close to projected performance, then they will have set new standards which the commercial vendors will find hard to ignore. However, to convert this success into a general purpose commercial product will depend on developments in software. It is therefore essential that the software development effort is not platform specific, and is common to clusters of workstations and MPP. If this goal is achieved then it will be simpler to harness the ever-growing hardware capability, and allow an enterprising company to develop a cost-effective teraflop scale MPP by the end of this decade. Otherwise, the investment for developing the software and for educating the user community will remain too high for a startup company.

## 7. Speculations and Conclusions

Let me conclude with a few remarks concerning first software and then hardware. It is absolutely essential to push for common software for both MPP and clusters of workstations. It is in the long term interest of all vendors to make a joint effort to provide mutually compatible software. In fact this should be a national strategy and given the highest priority by MPP vendors.

The most cost-effective migration path for existing serial codes to a parallel environment is via a small cluster of workstations as it avoids the initial expense of buying an MPP. If the software is common to both platforms then, the problems best suited for MPP can exploit that technology cost effectively and without interruption at the appropriate stage in code development.

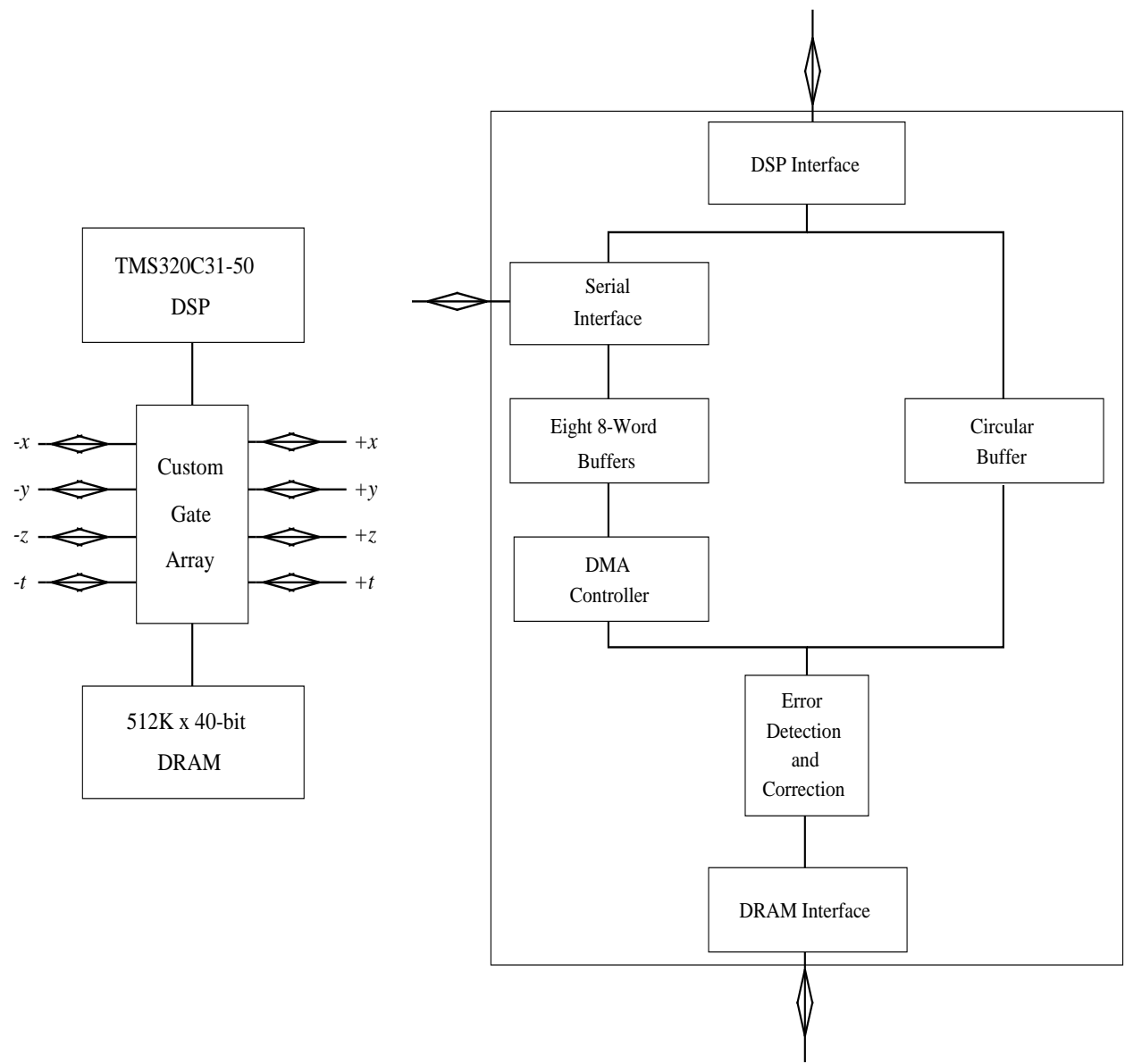

Fig.1 Schematic of (a) the node and (b) the custom gate array of the 0.5 Teraflop project.

Since this workshop is taking place so soon after the upheaval at Thinking Machines, it is difficult to avoid speculating on what the future of MPP would have been with them around. For those of us who have known many of the people involved in developing the Connection Machines and have worked closely with them, the breaking up of the company has been very disheartening. On a more impersonal level, it is a national tragedy that will set back parallel computing for some time to come. The biggest legacy of the pre-1994 Thinking Machines is the software. In my opinion, their CMF compiler is the best implementation of data parallel computing while their CMMD library is the cleanest and has the largest functionality for message passing codes. Even more importantly, they were close to having a synthesis of the two, global array declarations with facility for doing either message passing or data parallel computing. Their efforts towards this synthesis should be pursued vigorously as it fits into the long term goal of a common software. In terms of TMC's contribution to tools for program development, PRISM is by far the best debugger I have used. To summarize, I believe that, if TMC closes shop completely, then it is vital that every effort be made to preserve their software investment and to work towards making it available on a variety of platforms.

Concerning hardware there are two issues that I would like to stress. First, in all problems that I am familiar with, the data to be communicated is needed immediately afterwords for computations and does not need to be stored in memory first. It is therefore important that we figure out efficient ways of removing the unnecessary stores and loads. Second, the biggest bottleneck in obtaining high performance on today's MPP is moving data between memory (local or off node) and the register files. It would therefore help tremendously in the design and long term stability of optimizing compilers if memory access is pipelined and the size and characteristic of register files is standardized. Today's technology shows that high efficiency in loads and stores can be achieved with a pipe of length 8. In that case a register file that is at least 64 words long is needed to implement a complex $A + B \times C$ efficiently. Another very useful way to hide the communication latency is to allow for pre-fetches and stores in memory via DMA so that it can be done interrupting computations.

It is common knowledge amongst those who have simulated very large systems that it is very easy to compute for long periods of time but murder to store, restore, or frequently access large data files. I/O to disks and storage media is a major limiting factor today and will continue to be the Achilles heel as the disparity between FPU speed and data bandwidth to various peripherals grows. It seems like it will take a miracle to get these two aspects of computing balanced without developing equally efficient parallel I/O to disk and peripherals.

What can we look forward to? Pundits who study the computer market forecast that the entertainment industry will drive I/O capability, PC's and simple control processors will drive improvements in speed and packaging of components like memory and CPU, and signal processing will provide ways of increasing communication bandwidth. I would like to add that Grand Challenge problems should be used as the motivation for the design of the glue, *i.e.* the software paradigms and fast inter-node communications strategies to produce cost-effective MPP with teraflop and petaflop capabilities. I am totally convinced that parallel computing will be the platform on which all large-scale computations are done in the future. It is therefore my wish and sincere hope that commercial vendors will accept some common standards and strive towards a common software.

## Acknowledgements

I thank Ann Hayes and Margret Simmons for inviting me to participate in this workshop, and Andy White for prodding the large-scale users at LANL to think about the future of high performance computing. I also thank Rich Brower, Doug Kothe, Bob Malone, Bob Mawhinney, Manjeet Sahota, Doug Toussaint, A. Ukawa and Mike Warren for providing information about their projects, and Tanmoy Bhattacharya, Ralph Brickner and Wendy Schaffer for discussions. I gratefully acknowledge the tremendous support provided by the Advanced Computing Lab at Los Alamos. The simulations of QCD described here would not have been possible without the resources of the CM-5 at LANL and I thank DOE for a Grand Challenges allocation. The CDPEAC optimizations of LQCD were carried out by my collaborator Tanmoy Bhattacharya.